\documentclass{emulateapj} 

\usepackage{ulem}
\usepackage{color}

\usepackage{graphicx}

\usepackage[small]{caption}

\usepackage{amssymb,amsmath} 

\shorttitle{SS433's orbital eccentricity}
\shortauthors{Doolin \& Blundell}

\begin{document}

\title{The precession of SS433's radio ruff on long timescales}

\author{Samuel Doolin\altaffilmark{1} and Katherine M.\ Blundell\altaffilmark{1}}

\altaffiltext{1}{University of Oxford, Department of Physics, Keble
 Road, Oxford, OX1 3RH, U.K.}

\begin{abstract}

Roughly perpendicular to SS433's famous precessing jets is an outflowing ``ruff'' of radio-emitting plasma, revealed by direct imaging on milli-arcsecond scales. Over the last decade, images of the ruff reveal that its orientation changes over time with respect to a fixed sky co-ordinate grid. For example, during two months of daily observations with the VLBA by \citet{Mioduszewski:2004}, a steady rotation through $\sim10^{\circ}$ is observed whilst the jet angle changes by $\sim20^{\circ}$.

The ruff reorientation is \textit{not} coupled with the well-known precession of SS433's radio jets, as the ruff orientation varies across a range of $69^{\circ}$ whilst the jet angle varies across $40^{\circ}$, and on greatly differing and non-commensurate timescales. 

It has been proposed that the ruff is fed by SS433Õs circumbinary
disk, discovered by a sequence of optical spectroscopy by \citet{Blundell:2008}, and so we present the results of 3D numerical simulations of circumbinary orbits. These simulations show precession in the longitude of the ascending node of all inclined circumbinary orbits --- an effect which would be manifested as the observed ruff reorientation. Matching the rate of ruff precession is possible if circumbinary components are sufficiently close to the binary system, but only if the binary mass fraction is close to equality and the binary eccentricity is non-zero. 

\end{abstract}


\section{Introduction}

Imaging on milli-arcsecond scales is routine only at radio wavelengths with very long baseline interferometry (VLBI) techniques. Such techniques, applied to observations of the microquasar SS433 reveal extended emission roughly perpendicular to its famous precessing, knotty jets out to $\sim$100\,AU \citep{Paragi:1999,Paragi:2002,Blundell:2001,Mioduszewski:2004}, referred to as SS433's ruff. 

A succession of optical spectra of the Balmer H-$\alpha$ line revealed a pair of stationary components that were interpreted by \citet{Blundell:2008} as arising from a circumbinary disk (because of their lack of dependence on the orbital motion of the binary components).  \citet{Blundell:2008} further posit that this circumbinary disk feeds the radio ruff.

We have performed fully 3-D numerical simulations of circumbinary test particles in orbit around synthetic binary systems as a function of mass fraction and orbital eccentricity. Our simulations reveal clear trends in the behaviour of the plane of a circumbinary orbit with respect to the plane of the binary which we compare with time-resolved observations of SS433's ruff. 

\begin{figure}[!h]
\centering
\includegraphics[width=5.9cm]{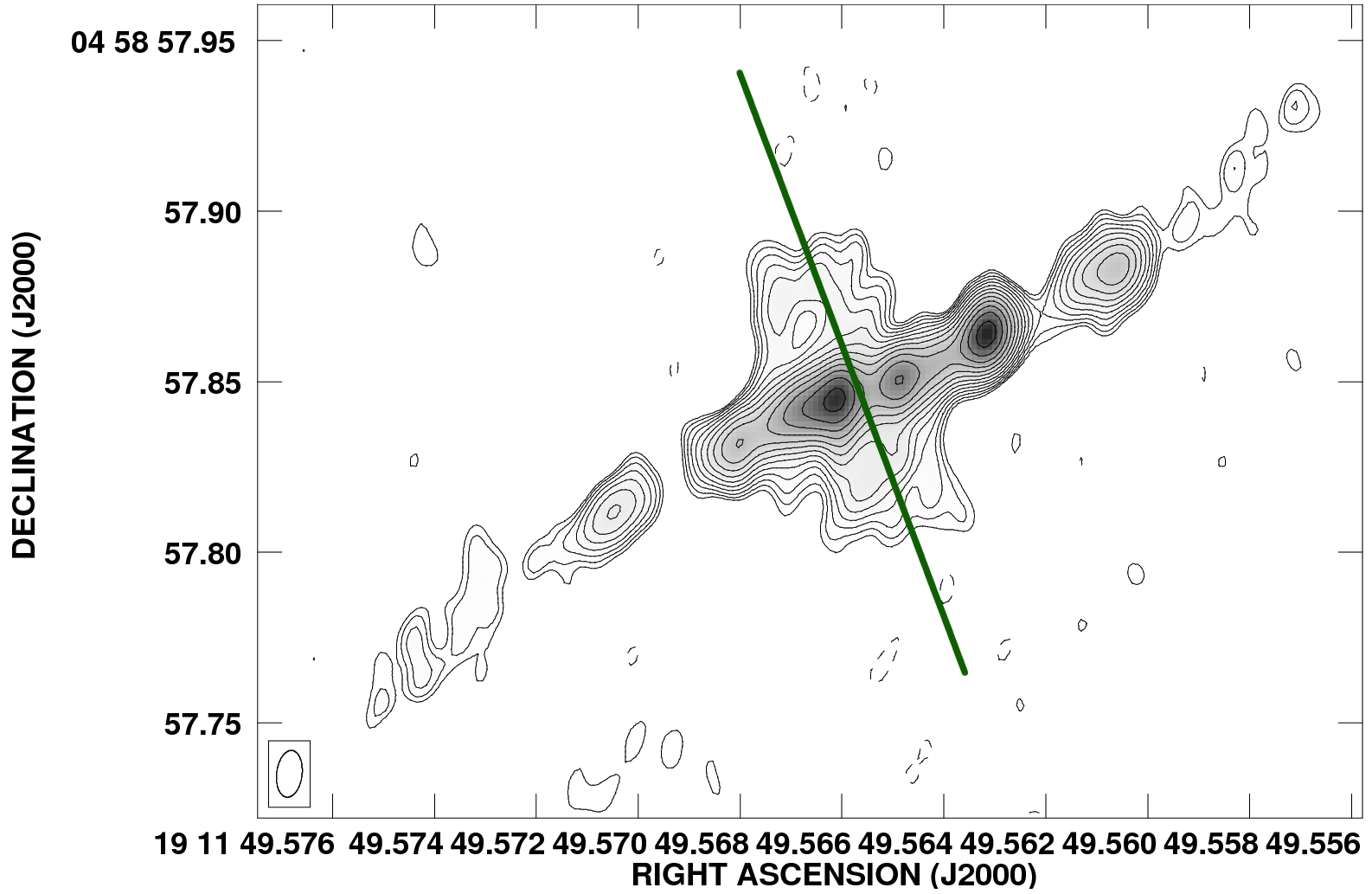}
\includegraphics[width=6.8cm]{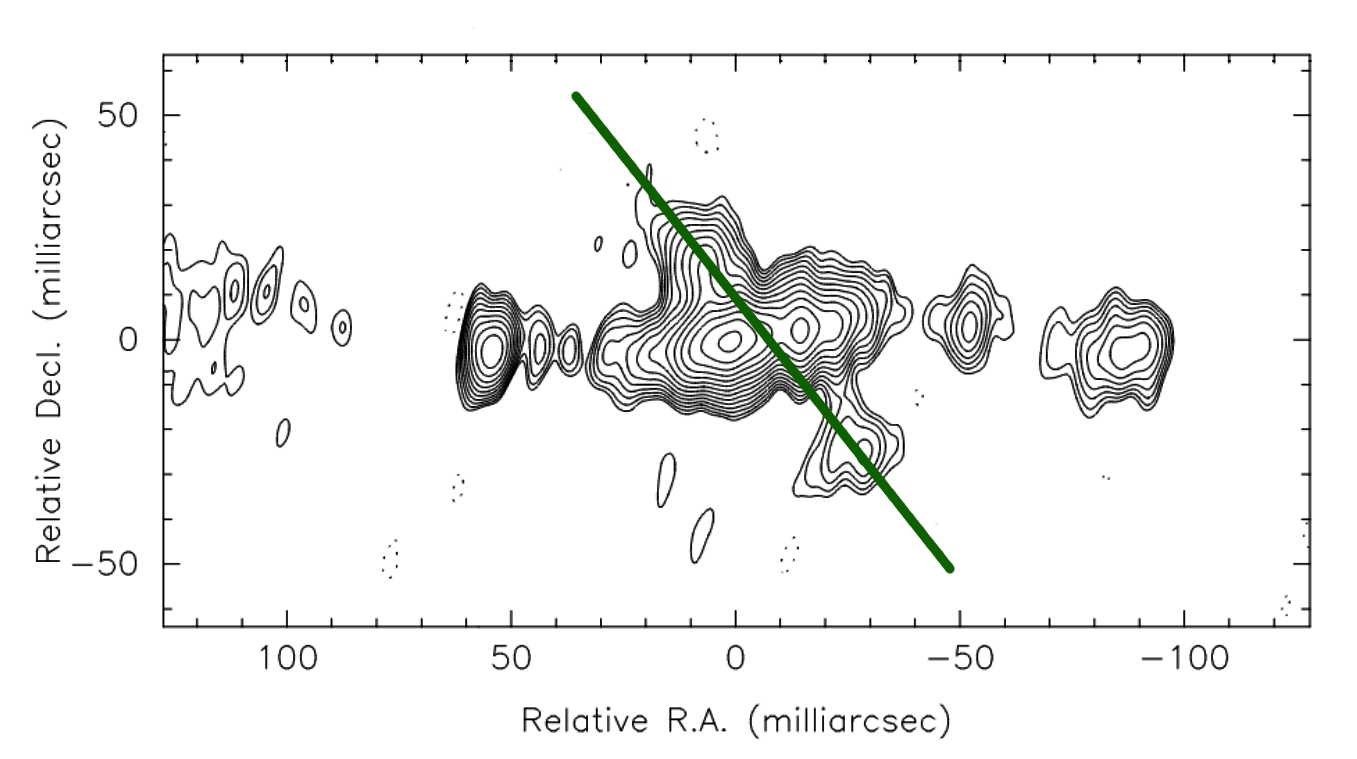}
\includegraphics[width=6.8cm]{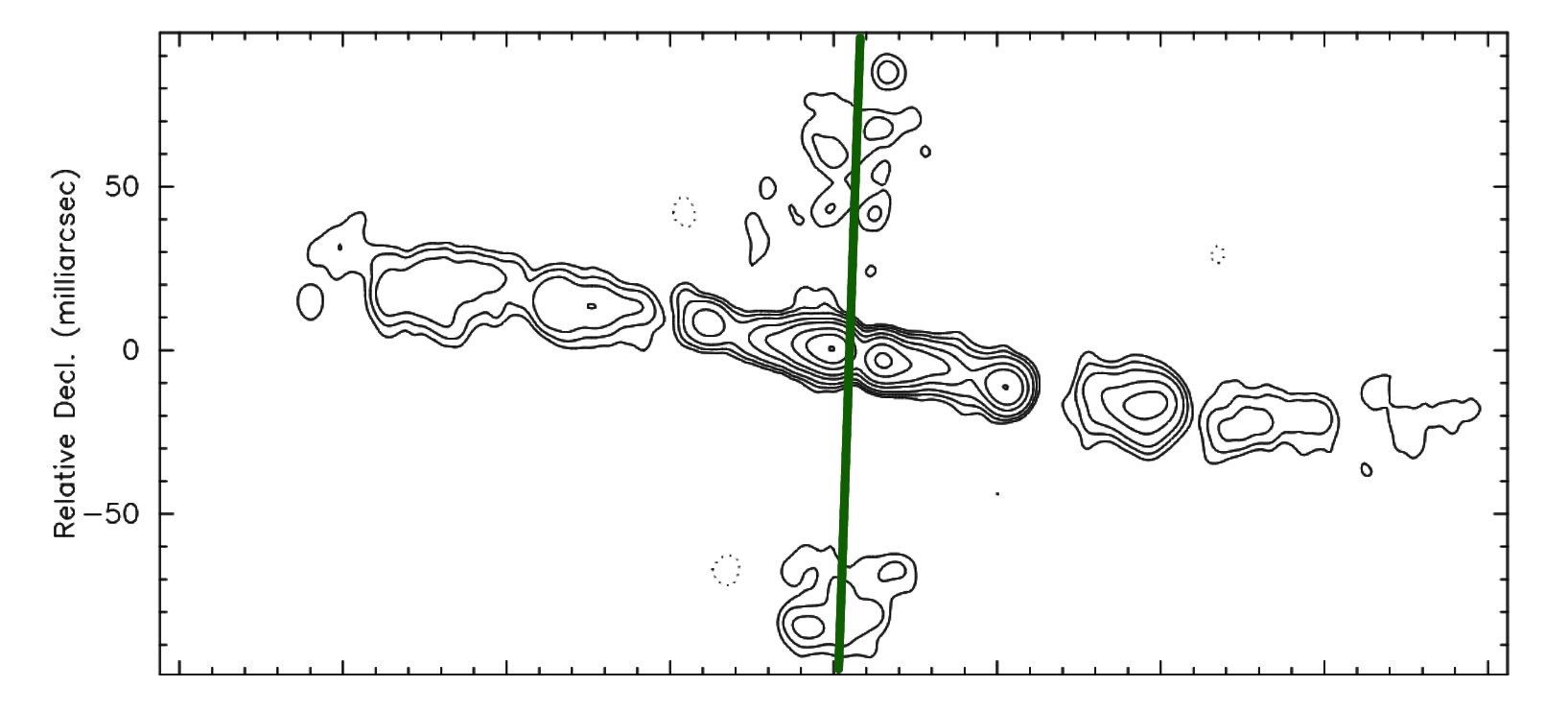}
\caption{Examples of radio images on milli-arcsec scales showing the famous precessing jets of SS433 and the \hbox{{\it differently}} precessing ruff -- highlighted in green. \\ {\it Top:} 7 March 1998, \citet{Blundell:2001} (their fig.\,2); \\ {\it Centre:} 6 June 1998, \citet{Paragi:2002a} (their fig 3); \\ {\it Lower:} 13 Feb 2000, \citet{Paragi:2002} (their fig 3a).
}
\label{fig:montage}
\end{figure}

\section{The changing orientation of SS433's ruff}
\label{sec:changing}

We extracted an estimate of the mean angle of the overall ruff from images presented by \citet{Paragi:1999,Blundell:2001,Paragi:2002,Paragi:2002a} and \citet{Mioduszewski:2004}, covering VLBI observations of SS433 from 1995 to 2003. We show in Figure~\ref{fig:montage} a subset of these images that clearly demonstrates the change in angle, projected on the plane of the sky, of SS433's ruff with time. 

From each of the VLBI images available to us we extracted this ``ruff angle'', measured anti-clockwise from celestial north. Where more than one published image was available for a given observation we measured all images. Higher resolution images reveal inner (more recently launched) ruff material, compared to the more extended ruff emission visible in lower-resolution images. 


We infer the epoch of emission (hereafter referred to as the ``corrected'' date) via the extent of the features of the observable ruff assuming a distance to SS433 of $5.5$ \rm{kpc} \citep{Blundell:2004} and the average wind speed of $10,000$\,\rm{km/s} as reported by \citet{Mioduszewski:2004}, although the launch mechanism for these observed high speeds remains to be established.   These corrections are of order $30$ days.    

\begin{figure}[!h]
	\centering
	\includegraphics[width=8.6cm]{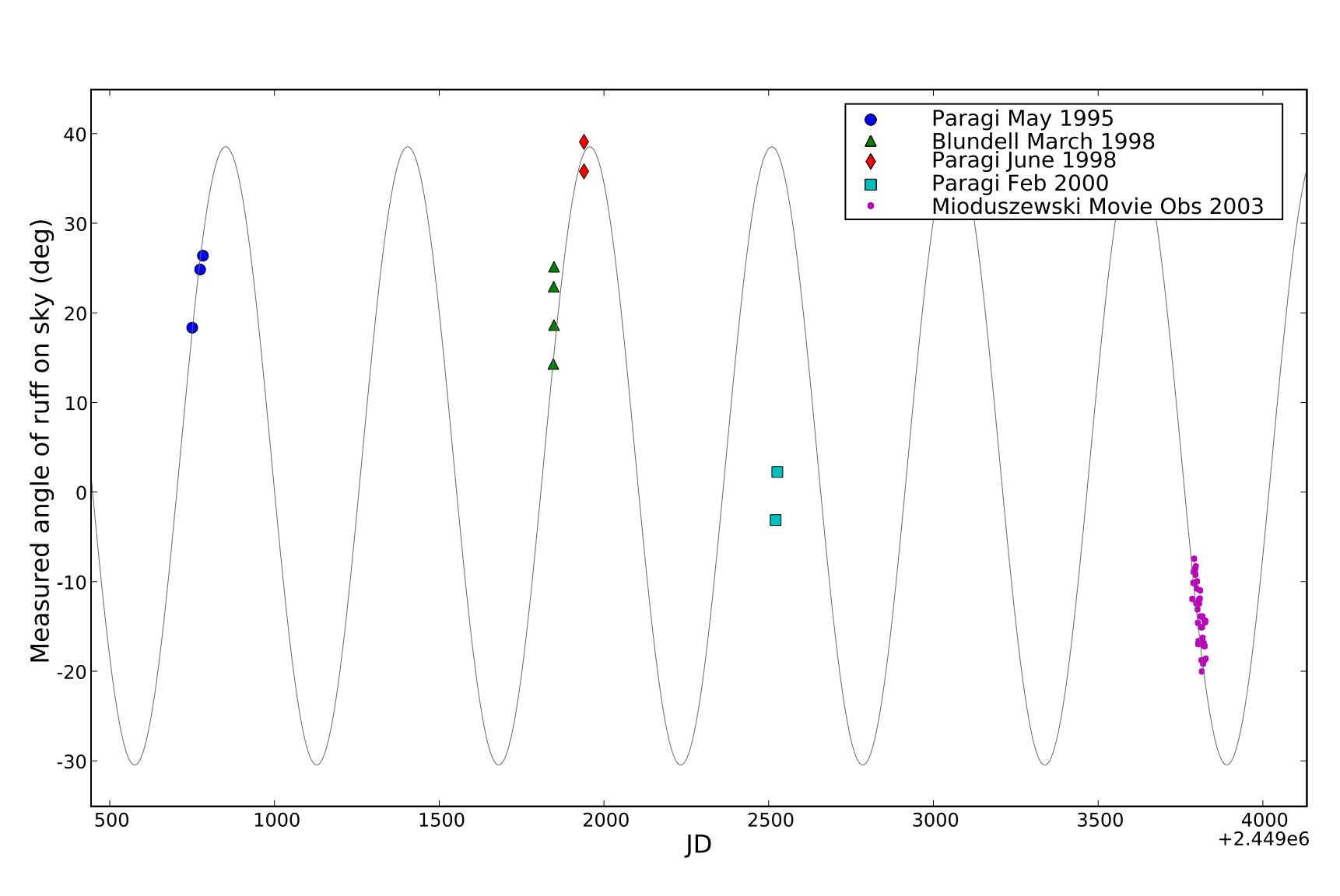}
	\caption{The data points are ruff angle vs corrected Julian Date (see \S\,\ref{sec:changing}), measured from all VLBI images of SS433 available to us. The sinusoidal line was fitted \textit{only} to the \citet{Mioduszewski:2004} data; details in text. }
	\label{angle_JD}
\end{figure}

In Fig \ref{angle_JD} we plot the ruff angle as a function of corrected Julian Date: these data clearly show systematic variations through a range of at least $50^{\circ}$. A fairly smooth, steady change is even revealed within the Mioduszewski movie observations of summer 2003. 

We also show in Figure \ref{angle_JD} the best fitting sinusoid. This fit is \textit{only} to the Mioduszewski data-points, giving an initial guess for the amplitude of $35^{\circ}$ with no vertical offset, after which frequency, phase, amplitude and vertical offset are fitted.

It is remarkable that many of the earlier data points lie (without any further fitting) on this sinusoid, which shows a periodicity in the apparent angle of the ruff of 552.5 days, or 42.2 binary orbital periods\footnote{One orbital period = 13.08 days \citep{kemp:1986}.}.

\section{Independence of ruff motion from other periodicities}
\label{periodicities}

We first consider whether the apparent precession of SS433's  ruff simply reflects other known periodicities in the SS433 system.  There are three principal periodicities to consider:  SS433's jet axis traces out a cone of semi-angle $\sim20^{\circ}$ every 162.375 days \citep{eikenberry:2001} known as its precession period; the orbital period of the binary is 13.08\,days \citep{kemp:1986}; and third, there is a nodding superimposed on the precession of the jet axis with a period of 6.06 days, manifested as beats with the precession period at 5.83 and 6.28\,days \citep{katz:1982}.

\begin{figure}[!h]
	\centering
	\includegraphics[width=9.5cm]{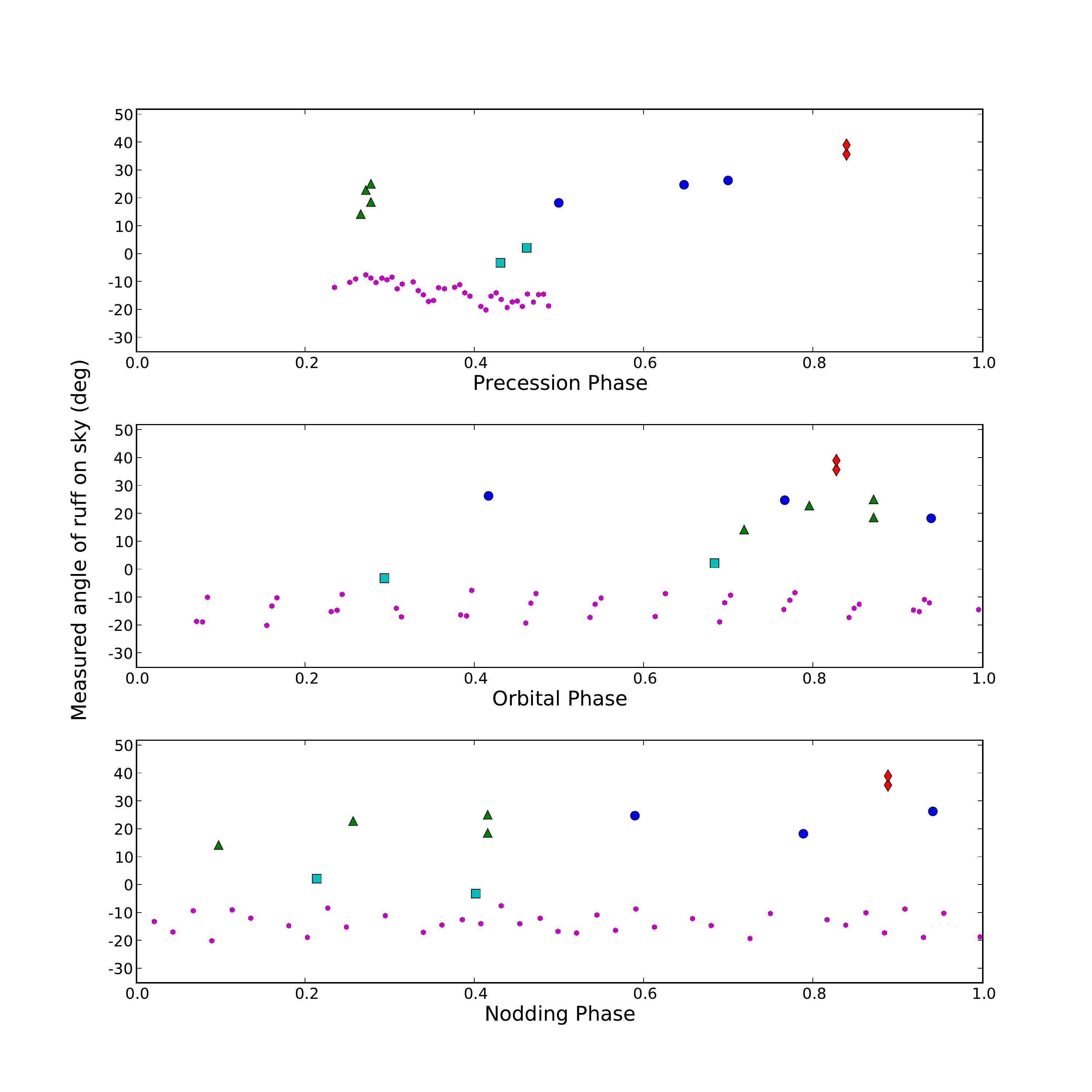}
	\caption{Plot of observed ruff angle versus precession, orbital and nodding phases: there is no evidence of any preferred ruff angle with any of these phases. Legend as Fig.~\ref{angle_JD}.  }
	\label{angle_phases}
\end{figure}

Fortunately, the time sampling of the available data enable us to rule out such correlations. In Fig\,\ref{angle_phases} we re-plot the measured ruff angles folded over SS433's three major periodicities and show an absence of a preferred ruff orientation with orbital, nodding or precession phase. We note that the periodicity plotted in Fig\,\ref{angle_JD} of 552.5 days is neither commensurate with the precession period of 162.375 days, orbital period of 13.08 days, nor nodding period of 6.06 days.

We also remark that the fit to the ruff angle varies through $69^{\circ}$, whereas the jet angle projected on the plane of the sky only has a range of $40^{\circ}$ \citep{Hjellming:1981}. 

\section{Description of the 3-D simulations}
\label{sims}

\citet{Blundell:2008} posit that SS433's ruff is fed from its circumbinary disk, and we further posit that the change in ruff angle arises because of an evolution in the orientation of this circumbinary structure. This motivated us to investigate the behaviour of circumbinary orbits.

We follow on from the 2D simulations of \citet{Holman:1999} by numerically integrating a suite of non-interacting test particles around a binary system of arbitrary eccentricity and mass ratio, in fully 3D calculations. 

We solve the orbit of the binary directly by applying a dual bisection and Newton-Raphson algorithm to Kepler's equation. We apply an adaptive step-size fourth and fifth-order Runge-Kutta integrator to integrate the system for $10^4$ binary orbital periods. During integration we remove test particles which come too close to either star, or escape the system.

\begin{figure}[!h]
	\centering
	\includegraphics[width=8.5cm]{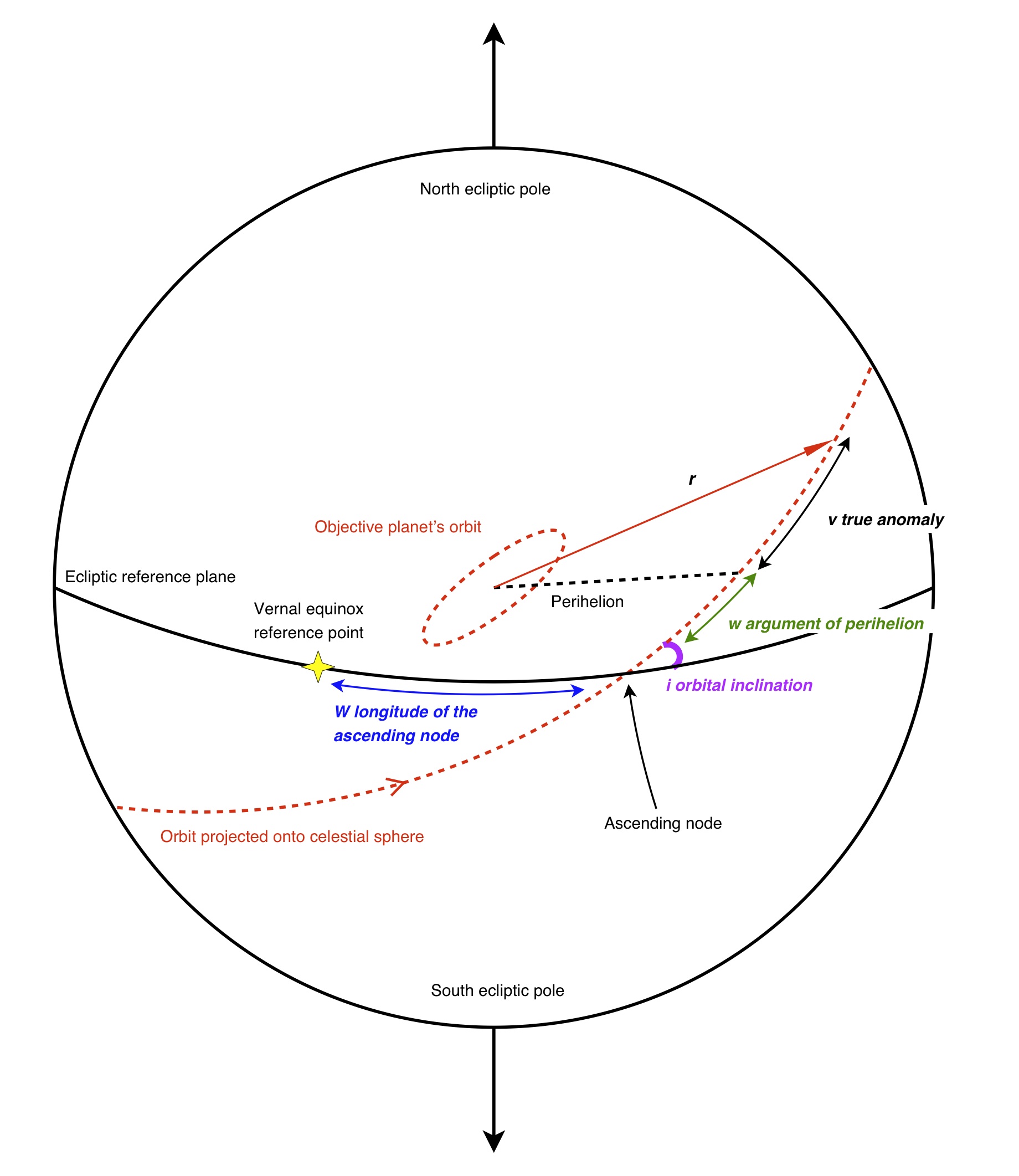}
	\caption{A representation of the Keplerian orbital elements}
	\label{fig-3D}
\end{figure}

A general orbit in 3D about a centre of mass may be described by the Keplerian orbital elements \mbox{($a$, $e$, $i$, $W$, $w$, $v$)} as illustrated by Figure \ref{fig-3D}. The eccentricity $e$, semi-major axis $a$ and true anomaly $v$ describe the motion of a test particle in its orbital plane, whilst the inclination $i$, longitude of the ascending node $W$ and argument of perihelion $w$ describe the orientation of the orbital plane of the test particles with respect to the plane of the binary. 

We track the orbital elements of each test particle about the centre of mass of the binary, outputting snapshots as a function of time to a MySQL database\footnote{These snapshots are of ``osculating" orbits: the Keplerian orbit about a central body that a test particle would have if other perturbations were not present.}. We deem an orbit to be \textit{stable} if it remains circular, but we notice that all inclined orbits precess in the longitude of the ascending node in a kozai-like manner \citep{Kozai:1962}; like a coin spinning on a tabletop (where the face of the coin is the plane of the circumbinary orbit). 

\section{Findings from our simulations}

Since SS433 is an eclipsing system we know that the orbital plane lies nearly perpendicular to the plane of the sky (\S\ref{sec-eclipsing}), so we set the plane of our simulated binary thus. We project the orbit of a test particle of inclination $i$ and longitude of the ascending node $W$ onto the plane of the sky and extract the angle that the orbit appears to subtend relative to the edge-on orbital plane. Hereafter this is referred to as the ``measured angle''.

If SS433's ruff is driven from circumbinary material \citep{Blundell:2008} then the changing angle presented by the ruff is manifested in our simulations as this ``measured angle''. As a test particle's orbital elements evolve over time it is the ``measured angle'' of its orbit that we contend is the ruff angle.

\begin{figure}[!h]
	\centering
	\includegraphics[width=\columnwidth]{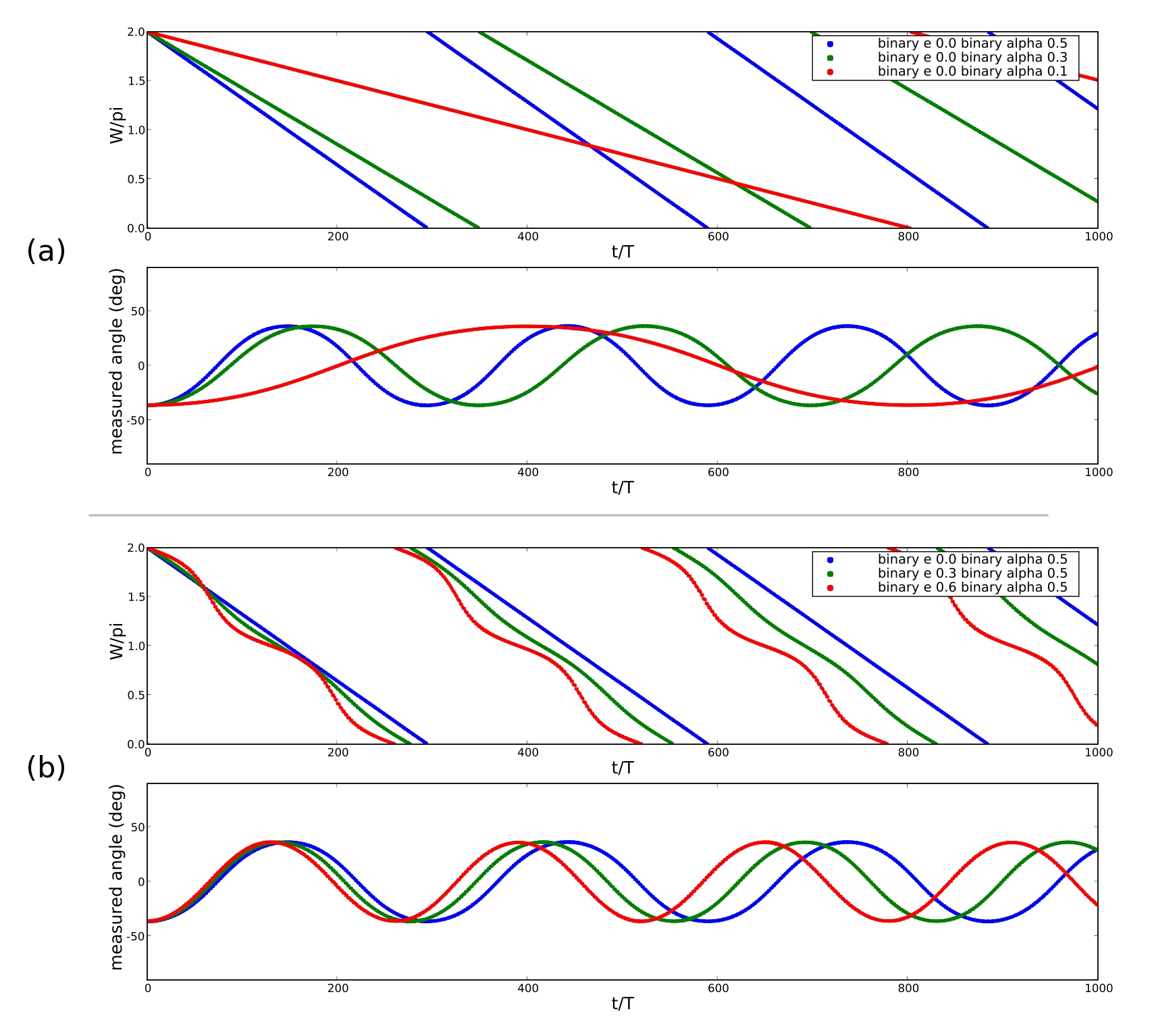}
	\caption{Tracking test particles: we plot the longitude of the ascending node and ``measured angle'' of a circumbinary orbit as a function of time, for example values of (a) binary mass fraction, and (b) binary eccentricity.}
	\label{fig-tracking}
\end{figure}

Figure \ref{fig-tracking} is a plot of the longitude of the ascending node and ``measured angle'' of an example test particle's orbit, as a function of binary eccentricity, binary mass fraction\footnote{Binary mass fraction $=M_1/(M_1+M_2)$, where $M_1$ is the lighter component and $M_2$ the heavier binary component.} (alpha) and time in units of binary period ($T$). This particle orbits at radius 3.0 times the semi-major axis of the binary system and has inclination angle $i = 36^{\circ}$ to the plane of the binary.

A full description of the parameter space explored by our simulations and the diagnostic machinery this provides for the interpretation of circumbinary orbits will be presented in a forthcoming paper. Here, we focus on the trends pertinent to the changing ruff angle in SS433:

\begin{itemize}
\item The closer the binary mass fraction to equality, the shorter the period of oscillation of the measured angle, i.e.\ precession of the longitude of the ascending node $W$.
\item The greater the binary eccentricity, the shorter the period of oscillation of the measured angle.
\item Fig\,\ref{fig-tracking}b also shows that increasing binary eccentricity produces a deviation away from a steady precession of the longitude of the ascending node $W$, and hence a departure from a pure sinusoidal curve in the oscillation of the measured angle. It is possible that the very small fraction of data points that do not lie on the sinusoidal fit of Fig.\,\ref{angle_JD} are deviant because SS433's binary orbit is eccentric. 
\item The closer the test particle orbits to the binary system, the shorter the period of oscillation of the measured angle. This of course arises because of the stronger coupling of angular-momentum exchange between closer-in particles and the binary, compared to particles that are further away. 
\end{itemize}

\subsection{Determination of orientations within SS433}
\label{sec:angles}

From the fit shown in figure~\ref{angle_JD} we deduce that the ruff precesses with an amplitude of $34.5^{\circ}$. At the maxima and minima of the curve the plane of the circumbinary disk contains our line of sight, actually perpendicular to the plane of the sky. Since the orbital plane is also into the plane of the sky we determine that the circumbinary disk is therefore inclined to the plane of the binary by $34.5^{\circ}$.  What determines this preferred inclination will be the subject of future investigation of the multiple components of this system.

Our fit also tells us the angle of rotation of the plane of the binary from north-south. Since we assume a symmetry in the precession of the ruff, spending equal time rotated clockwise and anticlockwise from the plane of the binary, the offset of the fit tells us that the plane of the binary (into the sky) is rotated $\sim2^{\circ}$ east of north. 

This angle is comparable with, but not equivalent to, the mean angle perpendicular to SS433's radio jets of $\sim10^{\circ}$ east of north \citep{Hjellming:1981,Blundell:2004}. But since the jets are produced by a warped and possibly non-co-planar accretion disk around the compact object we think this disagreement may be informing us about misalignment of the disk plane and jet-launch axis.

\section{Independent information on physical parameters}

\subsection{Independent constraints on SS433's orbital eccentricity} \label{sec-eclipsing}
\lastpagefootnotes

Optical photometry, such as in \citet{kemp:1986}, reveals that SS433's primary and secondary minima eclipses lie close to 0.5 of an orbital phase apart. This points towards a circular (non-eccentric) orbit, but is not conclusive because the orbit could still be eccentric if the axis of symmetry\footnote{binary orbital semi-major axis} is along our line-of-sight.

There are other pointers towards a non-zero eccentricity in SS433's orbit. First, the sinusoidally varying component of the jet speed \citep{Blundell:2005,Blundell:2007} depends on orbital phase, so there is something that breaks the symmetry of the orbit. Second, \citet{Perez:2009} find that the radius of the companion star must be rather large ($> 39$\,R$_{\odot}$) if the orbit is circular; but smaller radii for the star are permitted if the orbit has some eccentricity. 

\subsection{Independent confirmation of the mass fraction of SS433's binary components}

\citet{Blundell:2008}, from over a month of high resolution optical spectroscopy of the circumbinary disk, extracted the masses of SS433's binary components as being $\sim 16$ M$_{\odot}$ for the compact object and $\sim 24$ M$_{\odot}$ for the companion. Their measurements were also consistent with 18 and 22\,M$_{\odot}$ respectively, so on the basis of this independent determination a mass fraction of between $0.4$ and $0.45$ seems likely. 

We remark that the changing orientation of the precessing circumbinary disk will be manifested in how split the circumbinary lines appear to be \citep{Blundell:2008}. When the circumbinary disk is maximally face-on to Earth, then the line-splitting will be reduced (see \S\,\ref{sec:angles}), and with insufficient spectral resolution, becomes harder to resolve.

\section{Conclusions}

Comparing the results of our simulations (\S\ref{sims}) with observations of SS433's ruff angle (Fig~\ref{angle_JD}) we conclude that to achieve a short ruff precession period of $\sim 42$ binary periods it is necessary that the circumbinary material orbits very close in to the binary, i.e. a radius $\sim$ twice the binary semi-major axis.

In addition, it is necessary to have a high binary mass fraction, i.e. binary components of similar mass, and an eccentric binary orbit.

Our simulations show that inclined circumbinary orbits are surprisingly stable throughout binary mass-fraction/eccentricity parameter space. The motions of collective orbits, comprising circumbinary disks, have considerable potential to reveal system properties, if suitably time-resolved, high-resolution data are available. For the case of SS433, its changing ruff angle seems to confirm a mass fraction close to equality and a non-zero eccentricity. 

\acknowledgments SD thanks STFC for a studentship and KMB thanks the
Royal Society for a University Research Fellowship. They both thank Sebastian Perez M. for his proof reading and the referee for helpful comments on the manuscript.

\bibliographystyle{apj}

\begin{thebibliography}{15}

\expandafter\ifx\csname natexlab\endcsname\relax\def\natexlab#1{#1}\fi

\bibitem[{Blundell {et~al.}(2001)Blundell, Mioduszewski, Muxlow, Podsiadlowski,
  \& Rupen}]{Blundell:2001}
Blundell, K.~M., Mioduszewski, A., Muxlow, T., Podsiadlowski, P., \& Rupen, M.
  2001, \apjl, 562, L79

\bibitem[{Blundell \& {Bowler}(2004)}]{Blundell:2004}
Blundell, K.~M., \& {Bowler}, M.~G. 2004, \apjl, 616, L159

\bibitem[{Blundell \& {Bowler}(2005)}]{Blundell:2005}
---. 2005, \apjl, 622, L129

\bibitem[{Blundell {et~al.}(2007)Blundell, {Bowler}, \&
  {Schmidtobreick}}]{Blundell:2007}
Blundell, K.~M., {Bowler}, M.~G., \& {Schmidtobreick}, L. 2007, \aap, 474, 903

\bibitem[{Blundell {et~al.}(2008)Blundell, Bowler, \&
  Schmidtobreick}]{Blundell:2008}
Blundell, K.~M., Bowler, M.~G., \& Schmidtobreick, L. 2008, \apjl, 678, L47

\bibitem[{Eikenberry {et~al.}(2001)Eikenberry, {Cameron}, {Fierce}, {Kull},
  {Dror}, {Houck}, \& {Margon}}]{eikenberry:2001}
Eikenberry, S.~S., {Cameron}, P.~B., {Fierce}, B.~W., {Kull}, D.~M., {Dror},
  D.~H., {Houck}, J.~R., \& {Margon}, B. 2001, \apj, 561, 1027

\bibitem[{Hjellming \& {Johnston}(1981)}]{Hjellming:1981}
Hjellming, R.~M., \& {Johnston}, K.~J. 1981, \apjl, 246, L141

\bibitem[{Holman \& Wiegert(1999)}]{Holman:1999}
Holman, M.~J., \& Wiegert, P.~A. 1999, \apjl, 117, 621

\bibitem[{Katz {et~al.}(1982)Katz, {Anderson}, {Grandi}, \&
  {Margon}}]{katz:1982}
Katz, J.~I., {Anderson}, S.~F., {Grandi}, S.~A., \& {Margon}, B. 1982, \apj,
  260, 780

\bibitem[{Kemp {et~al.}(1986)Kemp, {Henson}, {Kraus}, {Carroll}, {Beardsley},
  {Takagishi}, {Jugaku}, {Matsuoka}, {Leibowitz}, {Mazeh}, \&
  {Mendelson}}]{kemp:1986}
Kemp, J.~C., {Henson}, G.~D., {Kraus}, D.~J., {Carroll}, L.~C., {Beardsley},
  I.~S., {Takagishi}, K., {Jugaku}, J., {Matsuoka}, M., {Leibowitz}, E.~M.,
  {Mazeh}, T., \& {Mendelson}, H. 1986, \apj, 305, 805

\bibitem[{Kozai(1962)}]{Kozai:1962}
Kozai, Y. 1962, \aj, 67, 591

\bibitem[{Mioduszewski {et~al.}(2004)Mioduszewski, {Rupen}, {Walker},
  {Schillemat}, \& {Taylor}}]{Mioduszewski:2004}
Mioduszewski, A.~J., {Rupen}, M.~P., {Walker}, R.~C., {Schillemat}, K.~M., \&
  {Taylor}, G.~B. 2004, in Bulletin of the American Astronomical Society,
  Vol.~36, 967--+

\bibitem[{Paragi {et~al.}(2002{\natexlab{a}})Paragi, {Fejes}, {Vermeulen},
  {Schilizzi}, {Spencer}, \& {Stirling}}]{Paragi:2002a}
Paragi, Z., {Fejes}, I., {Vermeulen}, R.~C., {Schilizzi}, R.~T., {Spencer},
  R.~E., \& {Stirling}, A.~M. 2002{\natexlab{a}}, in New Views on Microquasars,
  ed. P.~{Durouchoux}, Y.~{Fuchs}, \& J.~{Rodriguez}, 261--+

\bibitem[{Paragi {et~al.}(2002{\natexlab{b}})Paragi, {Fejes}, {Vermeulen},
  {Schilizzi}, {Spencer}, \& {Stirling}}]{Paragi:2002}
Paragi, Z., {Fejes}, I., {Vermeulen}, R.~C., {Schilizzi}, R.~T., {Spencer},
  R.~E., \& {Stirling}, A.~M. 2002{\natexlab{b}}, {in Proceedings of the 6th EVN Symposium}

\bibitem[{Paragi {et~al.}(1999)Paragi, {Vermeulen}, {Fejes}, {Schilizzi},
  {Spencer}, \& {Stirling}}]{Paragi:1999}
Paragi, Z., {Vermeulen}, R.~C., {Fejes}, I., {Schilizzi}, R.~T., {Spencer},
  R.~E., \& {Stirling}, A.~M. 1999, \aap, 348, 910

\bibitem[{Perez \& Blundell(2009)}]{Perez:2009}
Perez, S., \& Blundell, K. M. 2009, MNRAS in press

\end{thebibliography}

\end{document}